\documentclass[aps,prd,onecolumn,groupedaddress,showpacs,nofootinbib,amssymb]{revtex4}
\usepackage[T1]{fontenc}
\usepackage[latin1]{inputenc}
\usepackage{graphicx}
\usepackage[english]{babel}
\usepackage{amsmath}
\usepackage{amssymb}
\usepackage{amsfonts}

\begin{document}

\def\pp{{\, \mid \hskip -1.5mm =}}
\def\cL{{\cal L}}
\def\be{\begin{equation}}
\def\ee{\end{equation}}
\def\bea{\begin{eqnarray}}
\def\eea{\end{eqnarray}}
\def\beq{\begin{eqnarray}}
\def\eeq{\end{eqnarray}}
\def\tr{{\rm tr}\, }
\def\nn{\nonumber \\}
\def\e{{\rm e}}

\title{Accelerating cosmologies from non-local higher-derivative gravity}

\author{Salvatore Capozziello$^1$, Emilio Elizalde$^2$,
Shin'ichi Nojiri$^3$, Sergei D. Odintsov$^4$}

\affiliation{\it $^1$Dipartimento di Scienze fisiche, Universit\`a
di Napoli {}``Federico II'', INFN Sez. di Napoli, Compl. Univ. di
Monte S. Angelo, Edificio G, Via Cinthia, I-80126, Napoli, Italy\\
$^2$Consejo Superior de Investigaciones Cient\'\i ficas
ICE/CSIC-IEEC, Campus UAB, Facultat de Ci\`encies, Torre
C5-Parell-2a pl, E-08193 Bellaterra (Barcelona) Spain\\
$^3$ Department of Physics, Nagoya University, Nagoya 464-8602, Japan\\
$^4$Instituci\'o Catalana de Recerca i Estudis Avan\c{c}ats (ICREA) and
Institut de Ci\`encies de l'Espai (IEEC-CSIC), Campus UAB, Facultat
de Ci\`encies, Torre C5-Par-2a pl, E-08193 Bellaterra (Barcelona), Spain.}

\begin{abstract}

We study accelerating  cosmological solutions of a general class
of non-linear gravities which depend on Gauss-Bonnet and other higher
derivative invariants. To achieve this goal a local formulation with
auxiliary scalars for arbitrary higher-derivative non-local gravity is
developed. It is demonstrated that non-local Gauss-Bonnet gravity 
can be reduced, in the local formulation, to a model of string-inspired
scalar-Gauss-Bonnet gravity. A natural unification, in the theory here 
developed, of the early-time inflation epoch with a late-time 
acceleration stage can also be realized.

\end{abstract}


\maketitle

\noindent
{\bf 1. Introduction.}
It has been seen recently that not only early-time inflation but also
late-time acceleration may be naturally constructed in modified gravity
(for a review of modified gravities, see \cite{review}, and for review of
observational predictions from such theories, see \cite{capo}). It is
remarkable
that, moreover, a consistent unification of early-time inflation with
the late-time dark energy universe can indeed be achieved in modified
gravity (for a first such example, see
\cite{prd}). However, the models under consideration represent local
theories. At the same time, it is also well-known that quantum gravity and
stringy considerations lead to a low-energy effective action which
contains non-local, as well as local terms.

The proposal to use specific non-local gravity depending only on the scalar
curvature as a dark energy source has been made in Refs.~\cite{woodard,
Nojiri:2007uq}. Moreover, it has been demonstrated that different
cosmological solutions including phantom and/or quintessence
acceleration can be found, for specific versions of such non-local gravities
\cite{Nojiri:2007uq, koivisto, Jhingan:2008ym}. In addition, the
unification of early-time inflation with late-time acceleration in such
theory, which turns out to be consistent with local tests, is also possible
\cite{Nojiri:2007uq}. (A different approach to string-inspired non-local
cosmology has been discussed in Ref.~\cite{arefeva}). However, the
structure of such non-local gravity in the local formulation with
auxiliary fields is very complicated. In particular, an explicit de Sitter
solution has been only found for a very restricted (exponential) class of models.

In the present paper we consider a more general class of non-local gravities
which depend on Gauss-Bonnet and other higher-derivative invariants.
First, we formulate the local version of the theory with auxiliary fields
and show that almost all models of this sort admit the de Sitter cosmology
solution. It is quite interesting to observe that a relation with string-inspired
scalar-Gauss-Bonnet gravity (with possible account of higher-order string
corrections) is found. The emergence of inflationary or late-time
acceleration as well as the possibility to unify early-time inflation with
late-time acceleration will be here established.

\noindent
{\bf 2. The de Sitter universe from non-local ${\cal G}$-gravity.}
We now consider the following non-local model
\be
\label{nlGB1}
S=\int d^4 x \sqrt{-g} \left(\frac{R}{2\kappa^2}
 - \frac{\kappa^2}{2\alpha} {\cal G}\Box^{-1}{\cal G} + {\cal L}_{m} \right)\ .
\ee
Here ${\cal L}_m$ is the matter Lagrangian and ${\cal G}$
the Gauss-Bonnet invariant, namely
${\cal G} = R^2 - 4 R_{\mu\nu} R^{\mu\nu} + R_{\mu\nu\rho\sigma}R^{\mu\nu\rho\sigma}$.
By introducing the scalar field $\phi$, one may rewrite the action
(\ref{nlGB1}) in a local form:
\be
\label{nlGB3}
S=\int d^4 x \left(\frac{R}{2\kappa^2} - \frac{\alpha}{2\kappa^2}\partial_\mu\phi \partial^\mu \phi
+ \phi {\cal G} + {\cal L}_{m} \right)\ .
\ee
In fact, from the $\phi$-equation, we find $\phi = - \frac{\kappa^2}{\alpha} \Box^{-1} {\cal G}$.
By substituting this expression into (\ref{nlGB3}), we reobtain (\ref{nlGB1}).
It is remarkable that the above action belongs to a specific class of
string-inspired, scalar-Gauss-Bonnet gravity which is known to be
consistent with the local tests and may serve as a realistic candidate for dark
energy\cite{Nojiri:2005vv}. Unlike what happens in standard stringy compactifications,
the scalar-GB coupling is not an exponential scalar function. The scalar potential is zero,
as it should be in the low-energy, perturbative approach to the string effective
action.

Let us now study the de Sitter solutions which may describe late-time
universe acceleration in such non-local gravity.
By assuming that the FRW metric with flat spatial part and $\phi$  only
depend on the cosmological time, the  equations of motion are
\be
\label{nlGB5}
0= - \frac{3H^2}{\kappa^2} + \frac{\alpha}{2\kappa^2}{\dot\phi}^2 - 24\dot \phi H^3 + \rho_m \ ,\quad
0= - \frac{\alpha}{\kappa^2}\left(\ddot\phi + 3H \dot\phi\right)
+ 24 \left(\dot H H^2 + H^4 \right)\ .
\ee
Here $\rho_m$ is the energy density of matter and $H$ is the Hubble rate $H\equiv \dot a/a$.
When the matter contribution is neglected, assuming $\dot \phi$ and $H$ are constant
$\dot\phi=c_\phi$, $H=H_0$, one can solve Eqs.~(\ref{nlGB5}) as follows:
$c_\phi = \frac{8\kappa^2}{\alpha}H_0^3$, $H_0^4 = - \frac{3\alpha}{160\kappa^4}$.
Therefore, if $\alpha$ is allowed to be negative, there is a de Sitter universe
solution, namely
\be
\label{nlGB8}
H=H_0 = \left(  - \frac{3\alpha}{160\kappa^4} \right)^{1/4}\ .
\ee
On the other hand, a negative $\alpha$ means that the scalar field has a `wrong'
kinetic term, like a ghost or phantom.
As we will see later in (\ref{nlGBr14}), the de Sitter space solution corresponds
to the case $a\to \infty$ (late-time acceleration) when matter is included.
Even at the early universe, before the matter was created, we can use
this solution for the description of the inflationary epoch.

Note that the second equation (\ref{nlGB5}) yields
\be
\label{nlGBr1}
\left( -\frac{\alpha}{\kappa^2} \dot\phi + 8H^2 \right)a^3 = C\ .
\ee
Here $C$ is a constant. We should note that the action (\ref{nlGB3}) is invariant
under a constant shift of the scalar field $\phi$: $\phi\to \phi + c$ ($c$ is a constant)
since ${\cal G}$ is a total derivative. The quantity $C$ is the Noether charge corresponding to
the invariance. By using (\ref{nlGBr1}), we may delete $\dot\phi$ in the first equation 
in (\ref{nlGB5}) and obtain,
\be
\label{nlGBr2}
0= - \frac{3H^2}{\kappa^2} - \frac{160\kappa^2 H^6}{\alpha}
+ \frac{16\kappa^2 C H^3}{\alpha a^3} + \frac{\kappa^2 C^2}{\alpha a^6} + \rho_m\ .
\ee
When $\rho_m$ is given by the sum of the contributions from all forms of matter with constant EoS
parameter $w_i$, one gets $\rho_m(a) = \sum_i \rho_{0i} a^{-3 \left(w_i + 1\right)}$.
Then we can solve algebraically Eq.~(\ref{nlGBr2}) with respect to $H$ as a
function of $a$: $H={\cal H}(a)$. By the definition of $H$, the equation has the form as 
$\dot a = a {\cal H}(a)$. Then we find
$t=\int \frac{da}{a{\cal H}(a)}$, which gives, at least formally, the solution
of (\ref{nlGBr2}) and therefore the FRW universe.

As a simple case, we may consider the case $C=0$. Then Eq.~(\ref{nlGBr2})
reduces to the following form
\be
\label{nlGBr5}
0 = {\cal F}(x) \equiv x^3 + \frac{3\alpha}{160\kappa^4} x
 - \frac{\alpha \rho_m (a)}{160 \kappa^2}\ .
\ee
Here $x\equiv H^2$ and therefore $x\geq 0$. Since
${\cal F}'(x)=3 \left(x^2 + \frac{\alpha}{160\kappa^4}\right)$,
when $\alpha>0$, ${\cal F}(x)$ is a monotonically increasing function of $x$.
On the other hand, when $\alpha<0$, we find
\be
\label{nlGBr7}
{\cal F}'\left(x_\pm\right) = 0\ ,\quad
{\cal F}\left(x_\pm\right)= \mp \left( - \frac{\alpha}{160\kappa^4}\right)^{3/2}
 - \frac{\alpha \rho_m}{160\kappa^2}\ ,\quad x_\pm \equiv
\pm \left( - \frac{\alpha}{160\kappa^4}\right)^{1/2}\ .
\ee
Since
\be
\label{nlGBr8}
{\cal F}(0)= - \frac{\alpha \rho_m}{160 \kappa^2}\ ,
\ee
when $\alpha>0$, since ${\cal F}(0)<0$ and ${\cal F}(x)$ is a monotonically increasing function,
there is only one positive solution in (\ref{nlGBr5}):
\be
\label{nlGBr9}
H^2 = x = \alpha_+ + \alpha_-\ ,\quad \alpha_\pm^3 = \frac{\alpha \rho_m}{320\kappa^2}
\left( 1 \pm \sqrt{1 + \frac{\alpha}{40\kappa^8 \rho_m^2}}\right)\ .
\ee
We now consider the case that $\rho_m$ comes from a unique kind of
matter, with  constant EoS parameter $w$.
Then, when $a\to 0$, one obtains
$\alpha_+^3 \to \frac{\alpha \rho_0}{160 \kappa^2} a^{-3(1+w)}$ and $\quad \alpha_-\to 0$.
Therefore, $H\propto a^{-(1+w)/2}$ and we find $a \propto t^{2/(1+w)}$,
which could be compared with the usual FRW universe in Einsteinian gravity, where
$a\propto t^{2/3(1+w)}$.
On the other hand, when $a\to \infty$, Eq.~(\ref{nlGBr9}) implies that $H\propto a^{-3(1+w)/2}$
and therefore $a\propto t^{2/3(1+w)}$, which reproduces the result in Einstein's gravity.
This indicates that  such non-local term becomes less dominant as compared
with the Einstein-Hilbert term at the late universe epoch.

We may consider the case when $\alpha<0$. Since ${\cal F}(0)>0$ in this case,
if ${\cal F}\left(x_+\right)>0$, there is no positive solution for $x$. On the other hand,
if ${\cal F}\left(x_+\right)<0$, that is, when
\be
\label{nlGBr10}
 - \alpha > 40 \kappa^6 \rho_m^2 \ ,
\ee
there are two positive solutions and one negative solution in (\ref{nlGBr5}), namely
\be
\label{nlGBr11}
x = \tilde\alpha_+ + \tilde\alpha_-,\ \tilde\alpha_+ \xi + \tilde\alpha_- \xi^2,\
\tilde\alpha_+ \xi^2 + \tilde\alpha_- \xi,
\quad \tilde\alpha_\pm^3 = \frac{\alpha \rho_m}{320\kappa^2}
\left( 1 \pm i \sqrt{ - 1 - \frac{\alpha}{40\kappa^8 \rho_m^2}}\right)\ .
\ee
Eq.~(\ref{nlGBr10}) indicates that there is a lower bound in $a$ as
$a \geq a_{\rm min} \equiv \left( - \frac{\alpha}{40\kappa^4 \rho_0^2}\right)^{1/6(1+w)}$.
Here it is assumed that $w>-1$.
We should note that even if $a=a_{\rm min}$, $x$ and therefore $H$ do not vanish, which means that
the existence of $a_{\rm min}$ does not imply the bounce solution.
It is likely that a universe with $a=a_{\rm min}$ could appear, say, as a quantum correction.

When $a$ is large, we find that ${\cal F}(0)$ in (\ref{nlGBr8}) vanishes
and $\alpha_\pm$ tends to a constant:
$\alpha_\pm^3 \to \pm i \frac{1}{640 \kappa^6} \sqrt{ - \frac{ \alpha^3}{10}}$, which gives
\be
\label{nlGBr14}
H^2 = x = 0,\ \pm\sqrt{ - \frac{3\alpha}{160\kappa^4}}\ .
\ee
The positive solution of $x$ in (\ref{nlGBr14}) corresponds to the de Sitter solution
(\ref{nlGB8}).

When the Einstein-Hilbert term in (\ref{nlGB1}) or (\ref{nlGB3}) can be neglected, by assuming
\be
\label{nlGB9}
\phi=\frac{a_\phi}{t^2}\ ,\quad H = \frac{h_0}{t}\ \left(a=a_0 t^{h_0}\right)\ ,
\ee
Eqs.~(\ref{nlGB5}) reduce to the following algebraic equations
\be
\label{nlGB10}
0=a_\phi\left(\frac{\alpha}{\kappa^2}a_\phi + 24 h_0^3\right)\ ,\quad
0= \left(h_0 - 1\right)\left(\frac{\alpha}{\kappa^2}a_\phi + 4 h_0^3\right)\ .
\ee
The equations (\ref{nlGB10}) have a non-trivial solution ($a_\phi\neq 0$
and $h_0\neq 0$): $h_0=1$ and $a_\phi = - \frac{24\kappa^2}{\alpha}$.

We can now include matter. Then, the first equation in (\ref{nlGB10}) is modified as
\be
\label{nlGB12}
0= - \frac{3H^2}{\kappa^2} + \frac{\alpha}{2\kappa^2}{\dot\phi}^2 - 24\dot \phi H^3 + \rho_m\ .
\ee
When one may neglect the Einstein-Hilbert term in (\ref{nlGB1}) or
(\ref{nlGB3}) again, and if we assume
(\ref{nlGB9}), there is a non-trivial solution when $h_0 = \frac{4}{3(w+1)}$.
In fact, Eqs.~(\ref{nlGB5}) reduce to the following algebraic expressions
\be
\label{nlGB14}
0 = \frac{2\alpha}{\kappa^2}a_\phi^2 + \rho_0 a_0^{-3(w+1)}
+ 48 a_\phi \left(\frac{4}{3(1+w)}\right)^3\ ,
\quad
0= \left(\frac{4}{3(1+w)} - 1\right)\left(\frac{\alpha}{\kappa^2}a_\phi
+ 4\left(\frac{4}{3(1+w)}\right)^3 \right)\ .
\ee
If the matter is radiation with $w=1/3$, the second equation is satisfied
and the first equation acquires the following form:
$0 = \frac{2\alpha}{\kappa^2}a_\phi^2 + \rho_0 a_0^{-4} + 48 a_\phi$.
Then, if $\tilde D \equiv \left(24\right)^2 - \frac{2\alpha \rho_0}{\kappa^2 a_0^4} > 0$,
there are non-trivial solutions:
$a_\phi = \frac{\kappa^2}{2\alpha}\left(-24 \pm \sqrt{\tilde D}\right)$.
On the other hand, when $w\neq 1/3$, we find
$a_\phi = - \frac{4\kappa^2}{\alpha} \left(\frac{4}{3(1+w)}\right)^3$,
$\rho_0 a_0^{-4} = \frac{160 \kappa^2}{\alpha} \left(\frac{4}{3(1+w)}\right)^6$.
The second equation requires that $\alpha$ should be positive.
Thus, a phantom-like universe solution from purely non-local
gravity, without the $R$-term, has been found.

We now consider another version of  (\ref{nlGB1}), that is,
\be
\label{nlGBrrr1}
S=\int d^4 x \sqrt{-g} \left(\frac{R}{2\kappa^2}
 - \frac{\kappa^2}{2\alpha} \left(\frac{1}{\cal G}\right)
\Box^{-1}\left(\frac{1}{\cal G}\right) + {\cal L}_{m} \right)\ .
\ee
Here $\alpha$ is a dimensional constant.
Naively this action could be relevant in the late universe stage, where the curvature, and
therefore ${\cal G}$, is small ($1/{\cal G}$ is large) owing to the non-locality of
the action (\ref{nlGBrrr1}). However, there is a possibility that the action could be
relevant even for the early universe.
By introducing three scalar fields, $\phi$, $\xi$,
and $\eta$, one may rewrite the action (\ref{nlGBrrr1}) under the following form:
\be
\label{nlGBrrr2}
S=\int d^4 x \sqrt{-g} \left(\frac{R}{2\kappa^2}
+ \frac{\alpha}{2}\partial_\mu \phi \partial^\mu \phi
+ \frac{\alpha}{\kappa^7}\frac{\phi}{\eta}
+ \xi \left(\eta - {\cal G}\right)\right)\ .
\ee
Again, the action reminds of string-inspired gravity, including the first
Gauss-Bonnet correction. However, it may correspond to some complicated
compactification and/or to the contribution of non-perturbative terms, because of the
absence of kinetic terms for the two scalars and the appearance of a non-trivial multi-scalar
potential.

In the FRW universe, the action leads to the following equations
\be
\label{nlGBrrr3}
0 = - \frac{3}{\kappa^2}H^2 - \frac{\alpha}{2}{\dot\phi}^2 + 24 \dot\xi H^3\ ,\quad
0 = \alpha \left(\ddot\phi + 3H\dot\phi\right) + \frac{\alpha}{\kappa^7 \eta}\ ,\quad
0 = - \frac{\alpha}{\kappa^2}\frac{\phi}{\eta^2} + \xi\ ,\quad
0 = \eta - 24 \left(H^4 + \dot H H^2\right)\ .
\ee
By assuming $\phi = c_\phi t$, $\eta = \eta_0$, $\xi = \xi_0 t$, and $H=H_0$,
with constants $c_\phi$, $\eta_0$, $\xi_0$, and $H_0$, the equations in (\ref{nlGBrrr3}) reduce
to the algebraic ones
\be
\label{nlGBrrr5}
0 = - \frac{3}{\kappa^2}H_0^2 - \frac{\alpha}{2}c_\phi^2 + 24 \xi_0 H_0^3\ ,\quad
0 = 3 \alpha H_0 c_\phi + \frac{\alpha}{\kappa^7 \eta_0}\ ,\quad
0 = - \frac{\alpha}{\kappa^2}\frac{c_\phi}{\eta_0^2} + \xi_0\ ,\quad
0 = \eta_0 - 24 H_0^4 \ .
\ee
We can solve Eqs.~(\ref{nlGBrrr5}) with respect to $c_\phi$, $\xi_0$, and $\eta_0$
as follows: $c_\phi = - \frac{1}{72\kappa^7 H_0^5}$,
$\xi_0 = - \frac{\alpha}{3(24)^3 \kappa^{14} H_0^{13}}$,
$\eta_0 = 24 H_0^4$, and $H_0^{12} = - \frac{7}{54(24)^2 \kappa^{12}}$.
Therefore if $\alpha$ is negative, there could be a de Sitter universe solution, where $H$
is a constant:
$H_0 = \left(- \frac{7}{54(24)^2}\right)^{1/12} \frac{1}{\kappa}$.
As in the case of (\ref{nlGB8}), the de Sitter solution may both describe late time acceleration
or inflation in the early universe.
Hence, we have demonstrated that non-local $G$-gravities of different types may
lead to an alternative description of dark energy and/or inflationary
dynamics. Of course, for realistic description some version of the theory  should
be elaborated in more detail.

\noindent
{\bf 3. General non-linear higher-derivative gravity.}
We can now consider a generalization of (\ref{nlGB1}) which depends on all
three curvature tensors (the case of two different functions $F$ may be
considered by analogy)
\be
\label{nlGB19}
S=\int d^4 x \sqrt{-g} \left(\frac{R}{2\kappa^2}
 - \frac{1}{2\beta} F\left(R,R_{\mu\nu},R_{\mu\nu\sigma\rho}\right)\Box^{-1}
F\left(R,R_{\mu\nu},R_{\mu\nu\sigma\rho}\right)\right)\ .
\ee
Here $F\left(R,R_{\mu\nu},R_{\mu\nu\sigma\rho}\right)$ is a scalar function
of the scalar curvature $R$, the Ricci tensor $R_{\mu\nu}$, and the Riemann tensor.
Introducing an auxiliary field $\phi$ as in
(\ref{nlGB3}), the action (\ref{nlGB19}) can be rewritten in a local form
\be
\label{nlGB20}
S=\int d^4 x \sqrt{-g} \left(\frac{R}{2\kappa^2}
 - \frac{\beta}{2}\partial_\mu\phi \partial^\mu \phi
 - \phi F\left(R,R_{\mu\nu},R_{\mu\nu\sigma\rho}\right) \right)\ .
\ee
Even if the action includes higher powers of the inverse of $\Box$
(compare with \cite{Jhingan:2008ym}) as
\be
\label{nlGB19b}
S=\int d^4 x \sqrt{-g} \left(\frac{R}{2\kappa^2}
 - \sum_{n=1}^N \frac{1}{2\beta_n} F_n\left(R,R_{\mu\nu},R_{\mu\nu\sigma\rho}\right)\Box^{-n}
F_n \left(R,R_{\mu\nu},R_{\mu\nu\sigma\rho}\right)\right)\ ,
\ee
we can still rewrite the action (\ref{nlGB19b}) in a local form by introducing
auxiliary scalar fields (compare with the case of on-local $R$-gravity
\cite{Jhingan:2008ym}).

One may consider a more complicated generalization, as
\be
\label{nlGB21}
S=\int d^4 x \sqrt{-g} \left(\frac{R}{2\kappa^2}
 - F_1\left(R,R_{\mu\nu},R_{\mu\nu\sigma\rho}\right)\Box^{-1}
F_2\left(R,R_{\mu\nu},R_{\mu\nu\sigma\rho}\right)\right)\ .
\ee
Here we assume that $F_1$ is a function different from $F_2$. By properly multiplying
the Planck scale $M_{Pl}$, we may set the mass dimensions of $F_1$ and $F_2$ to be 3.
Then, the above action can be written in the following form:
\bea
\label{nlGB22}
&& S=\int d^4 x \sqrt{-g} \left(\frac{R}{2\kappa^2}
 - \frac{1}{2} \left(F_1\left(R,R_{\mu\nu},R_{\mu\nu\sigma\rho}\right)
+ F_2\left(R,R_{\mu\nu},R_{\mu\nu\sigma\rho}\right) \right)\Box^{-1}
\left(F_1\left(R,R_{\mu\nu},R_{\mu\nu\sigma\rho}\right)
+ F_2\left(R,R_{\mu\nu},R_{\mu\nu\sigma\rho}\right)\right) \right. \nn
&& \left. \qquad + \frac{1}{2} F_1\left(R,R_{\mu\nu},R_{\mu\nu\sigma\rho}\right)\Box^{-1}
F_1\left(R,R_{\mu\nu},R_{\mu\nu\sigma\rho}\right)
+ \frac{1}{2} F_2\left(R,R_{\mu\nu},R_{\mu\nu\sigma\rho}\right)\Box^{-1}
F_2\left(R,R_{\mu\nu},R_{\mu\nu\sigma\rho}\right)\right) \ .
\eea
By introducing now three scalar fields, $\phi_1$, $\phi_2$, $\phi_3$,
we can rewrite (\ref{nlGB22}) in a local form as
\bea
\label{nlGB23}
S&=&\int d^4 x \left(\frac{R}{2\kappa^2} + \frac{1}{2}\partial_\mu\phi_1 \partial^\mu \phi_1
+ \frac{1}{2}\partial_\mu\phi_2 \partial^\mu \phi_2 \right. \nn
&& \left.  - \frac{1}{2}\partial_\mu\phi_3 \partial^\mu \phi_3
 - \left(\phi_1 + \phi_3\right) F_1\left(R,R_{\mu\nu},R_{\mu\nu\sigma\rho}\right)
 - \left(\phi_2 + \phi_3\right) F_2\left(R,R_{\mu\nu},R_{\mu\nu\sigma\rho}\right) \right)\ .
\eea
Note that if $F_1,F_2$ include Gauss-Bonnet and higher-order terms which
are cubic and fourth-order in the curvature invariants, the above theory
represents the low-energy string effective action with higher-order
corrections. The de Sitter solutions for such effective action have been
studied in \cite{hd}.

As an example of (\ref{nlGB19}) or (\ref{nlGB20}), we may consider the case
$F\left(R,R_{\mu\nu},R_{\mu\nu\sigma\rho}\right)
= c_1 {\cal G} + c_2 R$, with constants $c_1$ and $c_2$.
Then, in the FRW metric, we find the following equations:
\bea
\label{nlGB25}
0 &=&  -\beta \left(\ddot\phi + 3H\dot\phi\right) - 24 c_1\left(\dot H H^2 + H^4\right)
 - 6 c_2 \left(\dot H + 2 H^2\right)\ ,\\
\label{nlGB26}
0 &=& - \frac{3H^2}{\kappa^2} + \frac{\beta}{2} + 24 c_1 \dot\phi H^3
 - 3c_2\left(\phi H^2 - \dot\phi H\right) + \rho_m\ .
\eea
When the contribution from matter can be neglected, that is, provided $\rho_m=0$, if $c_2/c_1$
is negative there is a de Sitter space solution:
$H = \sqrt{- \frac{c_2}{2c_1}}$ and $\phi = - \frac{1}{c_2\kappa^2}$.
As we have neglected the contribution from matter, again, the de Sitter solution could be
valid to describe late time acceleration or inflation before matter was created in the
very early universe.

Now, the following model can be considered:
\be
\label{nlGB28}
S=\int d^4 x \sqrt{-g}\left\{ \frac{R}{2\kappa^2} + \left(c_1 {\cal G} + c_2 R \right)
f\left(\Box^{-1}\left(c_1 {\cal G} + c_2 R \right)\right) \right\}\ .
\ee
Here $c_1$ and $c_2$ are constant.
By introducing two scalar fields, $\xi$ and $\phi$, we may rewrite the action (\ref{nlGB28})
in a local form
\be
\label{nlGB29}
S=\int d^4 x \sqrt{-g}\left\{ \frac{R}{2\kappa^2} - \left(\xi - f(\phi)\right)
\left(c_1 {\cal G} + c_2 R \right) - \partial_\mu \xi \partial^\mu \phi \right\}\ .
\ee
One may check that the action (\ref{nlGB29}) is equivalent to the action
(\ref{nlGB28}).
By variation of $\xi$ and $\phi$, we have
\bea
\label{nlGB30}
&& 0=\Box \phi - \left(c_1 {\cal G} + c_2 R \right) \ ,\\
\label{nlGB31}
&& 0=\Box \xi + f'(\phi) \left(c_1 {\cal G} + c_2 R \right) \ .
\eea
If we delete $\phi$ in (\ref{nlGB29}) by using (\ref{nlGB30}), we reobtain the action
(\ref{nlGB28}).
In the FRW background, Eqs.~(\ref{nlGB30}) and (\ref{nlGB31}) can be rewritten as
\bea
\label{nlGB32}
&& 0 = - \left(\ddot \phi + 3H \dot \phi \right)
 - \left\{ 24 c_1 \left(\dot H H^2 + H^4 \right)
+ 6 c_2 \left(\dot H + 2 H^2 \right)\right\}\ ,\\
\label{nlGB33}
&& 0 = - \left(\ddot \xi + 3H \dot \xi \right)
+ f'(\phi)\left\{ 24 c_1 \left(\dot H H^2 + H^4 \right)
+ 6 c_2 \left(\dot H + 2 H^2 \right)\right\}\ .
\eea
On the other hand, the equation corresponding to the FRW one has the following form
\be
\label{nlGB34}
0 = - \frac{3}{\kappa^2} H^2 + 24 c_1 \left( \dot \xi - f'(\phi) \dot \phi \right) H^3
+ 3 c_2 \left\{ \left(\xi - f(\phi) \right) H^2
 - \left( \dot \xi - f'(\phi) \dot \phi \right) H\right\} + \rho_m \ .
\ee
We now investigate if there could be a de Sitter space solution. We assume $H$ is a constant
$H=H_0$ and neglect matter, $\rho_m = 0$. Then, Eq.~(\ref{nlGB32}) can be integrated to be
\be
\label{nlGB35}
\phi = - \left( 8 c_1 H_0^3 + 4 c_2 H_0 \right) t - \phi_1 \e^{-3 H_0 t} + \phi_2 \ .
\ee
Here $\phi_1$ and $\phi_2$ are constants of integration.
For simplicity, we consider the case $\phi_1 = \phi_2 = 0$ and we assume that $f(\phi)$ has the
following form:
\be
\label{nlGB36}
f(\phi) = f_0 \e^{b\phi}\ .
\ee
Here $b$ is a constant. By substituting (\ref{nlGB35}) with $\phi_1 = \phi_2 = 0$ into
(\ref{nlGB36}) and integrating (\ref{nlGB33}), we obtain
\be
\label{nlGB37}
\xi = - \frac{3 H_0 f_0}{3H_0 - \beta} \e^{-\beta t} - \frac{\xi_1}{3 H_0}\e^{-3 H_0 t}
+ \xi_2\ .
\ee
Here $\xi_1$ and $\xi_2$ are constants of integration and $\beta$ is defined by
\be
\label{nlGB38}
\beta \equiv \left( 8 c_1 H_0^3 + 4 c_2 H_0 \right)b\ .
\ee
Substituting now (\ref{nlGB35}) with $\phi_1 = \phi_2 = 0$
and (\ref{nlGB38}) into (\ref{nlGB34}), one gets
\be
\label{nlGB39}
0 = - 3H_0^2 \left(\frac{1}{\kappa^2} + c_2 \xi_2\right)
+ \frac{\left(24 c_1 H_0^3 + 3c_2 H_0\right)\beta^2
 -  3 c_2 H_0^2 \beta }{3H_0 - \beta} f_0 \e^{-\beta t}
+ \left(24 c_1 H_0^3 - 4 c_2 H_0 \right) \xi_1\e^{-3H_0 t}\ .
\ee
When $\beta\neq 3H_0$ nor $\beta\neq 0$, Eq.~(\ref{nlGB39}) yields
\be
\label{nlGB40}
\xi_2 = - \frac{1}{c_2 \kappa^2}\ ,
\ee
and
\be
\label{nlGB41}
0 = 6 c_1 H_0^2 - c_2\quad \mbox{or}\quad \xi_1 = 0\ .
\ee
Furthermore,
\be
\label{nlGB42}
\beta = -1 - \frac{8c_1H_0}{c_2}\ \mbox{or}\ b = - \frac{1 + \frac{8c_1H_0}{c_2}}{8c_1 + 4c_2 H_0}\ .
\ee
Then if we choose $b$ in (\ref{nlGB36}) as in (\ref{nlGB42}), there is a de Sitter
solution with $H=H_0$.
When $c_1=0$, that is, without the Gauss-Bonnet term, we have $\xi_1=0$ and (\ref{nlGB40}).
Eq.~(\ref{nlGB39}) also gives $\beta = -1$,
which corresponds to the solution in \cite{Nojiri:2007uq}.
In the same way as in Ref.~\cite{Nojiri:2007uq} one can suggest the
unification of early-time inflation with late-time acceleration within
the theory. It is also remarkable that unlike in the case of
non-local gravities with $R$-dependence only, the above theory is very rich and
does admit de Sitter universe solutions for a variety of actions.

The unification of late-time acceleration and inflation in the early
universe can in fact be easily achieved  by adding a
$F(R)$-term to the action of non-local Gauss-Bonnet gravity (\ref{nlGB1}), as
\be
\label{nlGB45}
S=\int d^4 x \sqrt{-g} \left(\frac{R}{2\kappa^2}
 - \frac{\kappa^2}{2\alpha} {\cal G}\Box^{-1}{\cal G} + F(R) + {\cal L}_{m} \right)\ ,
\ee
or to the action
\be
\label{nlGB46}
S=\int d^4 x \left(\frac{R}{2\kappa^2} - \frac{\alpha}{2\kappa^2}\partial_\mu\phi \partial^\mu \phi
+ \phi {\cal G} + F(R) + {\cal L}_{m} \right)\ .
\ee
Here $F(R)$ is a proper function of the scalar curvature $R$ (the above 
theory has been proposed for dark energy description in Ref.~\cite{petr}). Then the first equation in
Eq.~(\ref{nlGB5}) is changed as
\be
\label{nlGB47}
0= - \frac{3H^2}{\kappa^2} + \frac{\alpha}{2\kappa^2}{\dot\phi}^2 - 24\dot \phi H^3
 - F(R) + 6\left(H^2 + \dot H\right) F'(R) - 36\left(4H^2 \dot H + H \ddot H\right)F''(R)
+ \rho_m \ ,
\ee
but the second equation is not changed and therefore we have (\ref{nlGBr1}) again.

One can consider the case where the early inflation could be mainly
generated by the
$F(R)$-term and the late-time accelerating expansion could be generated by the non-local
Gauss-Bonnet term.
We now choose $F(R)= \beta R^2$.
Here $\beta$ is a constant. We choose $\alpha$ to be very small
and $\phi$ starts with $\phi=0$. Hence, at the early universe the non-local Gauss-Bonnet
term is very small and could be neglected. We also assume that the contribution from matter
could be also neglected in the early universe. Then due to the $F(R)$-term,
there occurs $R^2$-inflation. After the end of inflation,
the radiation/matter dominance era takes over, where $H$ behaves
as $H\sim h_0/\left(t+t_0\right)$. Then by integrating (\ref{nlGBr1}), we find
\be
\label{nlGB48}
\phi = \phi_0 + \frac{\kappa^2}{\alpha} \left(- \frac{8h_0^2}{t+t_0}
 - \tilde C \left(t+t_0\right)^{-3h_0 +1} \right)\ .
\ee
Here $\phi_0$ and $\tilde C$ are constants of integration, which could be determined
by proper initial conditions. When $t$ becomes large, the value of $\phi$ goes to
$\phi_0$. If $\phi_0$ can be large enough, the non-local Gauss-Bonnet term can be
dominated and the universe goes to asymptotically de Sitter space, which describes
late-time acceleration. Thus, a mixture of non-local Gauss-Bonnet gravity
with local modified gravity, on top of being a rather natural theory made up
 of the fundamental ingredients one has at hand, yields a very realistic model for
 the unification of early-time inflation with late-time acceleration.

\noindent
{\bf 4.Discussion.}

In summary, we have demonstrated in this paper that a large class of
non-local higher-derivative
gravities admits accelerating cosmologies which can naturally describe
inflation, dark energy and/or the wholly unified inflation-radiation/matter
dominance-dark energy, in the universe expansion history. We have chiefly
concentrated on models which depend on a Gauss-Bonnet invariant but this
was essentially for technical reasons: the simplification of the equations of motion.
There is no problem to extend the study to account for
Riemann squared tensor- and/or Ricci squared tensor-dependent non-local
gravity. In addition, non-linear GB gravity can be represented as a specific
string-inspired gravity, what hints towards a fundamental origin from string theory.
Unlike in the case of pure curvature-dependent non-linear gravity, many different
versions of the theory under considerations exhibit de Sitter universe solutions
which maybe used to describe the different accelerating epochs. It is known that
such solutions represent the fixed points of the cosmological dynamics
where the effective equation of state parameter is $w=-1$. Moreover, the
 stability of the fixed points can be studied analytically or numerically.
Hence, it is
straightforward to construct the effective phantom and quintessence-like
cosmologies which lie below or above the corresponding de Sitter fixed points in
a similar way as in Ref.~\cite{Jhingan:2008ym}.

In order to select the most realistic version from the above theories, 
local tests, PPN analysis and accurate observational data must be used.
For instance, cosmologies derived from non-local higher-derivative gravity
should be experimentally tested in order
 to make a short list of most viable theories. From a genuine
cosmological viewpoint, several non-local models
give rise to very similar dynamics. A way to overcome such a shortcoming
could be to search for signatures in the stochastic background of
 gravitational waves in order to discriminate and constrain  models.
In fact, the production of primordial
gravitational waves, if revealed, is a robust prediction
 for any model attempting to describe the cosmological evolution at early epochs.
 It may be directly derived from the inflationary
scenario \cite{Watson}, which fits the WMAP data well, being the agreement 
particularly good with  exponential inflation and the spectral index $\simeq 1$.
The main characteristics of the gravitational background produced
by cosmological sources depend both on the emission properties of
each single source and on the source rate evolution with the redshift.
To this purpose,  one can  take into account the primordial
physical process which gave rise to a characteristic spectrum
$\Omega_{sgw}$ for the early stochastic background of relic scalar
gravitational waves by which we can recast the further degrees of freedom coming
from non-local higher derivative gravity. This approach
 can greatly contribute to
constrain the viable cosmological models. The physical process related
to the production has been analyzed for the first several tensorial
components of standard General Relativity. Actually the
process can be improved considering also the further scalar-tensor
components strictly related to the   degrees of
freedom of non-local gravity. It should emphasized that
the stochastic background of scalar gravitational waves can be described
 in terms of  scalar fields $\phi$ and characterized by a dimensionless
spectrum. In our case, considering a generic non-local theory 
of order $(2n+2)$ as, for example,
\be
{\cal L}=\sqrt{-g}F(R,\Box R,...\Box^k R, \Box^n R)\,,
\ee
it is straightforward to show that, by linearizing the theory, one obtains
\be
\prod_{k=0}^{n} (\Box-m_k^2) \phi_k=0\,,
\ee
with $n$ massive scalar fields ($m_k$ real) or tachyons ($m_k$ imaginary)
 conformally related to the curvature invariants.  The interferometric space
 experiment LISA could put tight constraints on such masses. An analysis
similar to the one in \cite{PPN} will be developed for the theory under 
consideration elsewhere.

\noindent
{\bf Acknowledgments}
This research was supported by a INFN-CICYT bilateral project, by {\it
Azione Integrata Italia-Spagna 2007} (MIUR Prot. No. 464,
13/3/2006) grant, and by MCIN (Spain) projects FIS2006-02842 and
PIE2007-50I023. The work by S.N. is supported by Min. of
Education, Science, Sports and Culture of Japan under grant no.
18549001 and Global
COE Program of Nagoya University provided by the Japan Society
for the Promotion of Science (G07).

\end{document}